\documentstyle[12pt,aasms4]{article}

\received{November 3 1999}
\accepted{February 23 2000}

\begin{document}

\title{The $L_x$-$T$, $L_x$-$\sigma$ and 
$\sigma$-$T$ relations for groups and clusters of galaxies}

\author{Yan-Jie Xue and Xiang-Ping Wu}

\affil{Beijing Astronomical Observatory and 
       National Astronomical Observatories, 
       Chinese Academy of Sciences, Beijing 100012, China}

\begin{abstract}
While in the hierarchical model of structure formation,
groups of galaxies are believed to be the scaled-down version of clusters
of galaxies, a similarity breaking in the fundamental laws 
may occur on the  group scale, 
reflecting a transition between galaxy-dominated and intracluster
medium dominated properties. In this paper, 
we present an extensive study of 
the relations between the X-ray luminosity ($L_x$), 
the temperature ($T$) of hot diffuse gas and the velocity 
dispersion ($\sigma$) of galaxies for groups and clusters of galaxies, 
based on the largest sample of 66 groups and 274 clusters drawn from 
literature. Our best fit $L_x$-$T$ and $L_x$-$\sigma$ relations for groups
read $L_x\propto T^{5.57\pm1.79}\propto \sigma^{2.35\pm0.21}$, which
deviates remarkably from those for clusters: 
$L_x\propto T^{2.79\pm0.08}\propto \sigma^{5.30\pm0.21}$.
The significance of these correlations have been justified
by both the co-consistency test and the Kendall's $\tau$ statistics.   
We have thus confirmed the existence of 
similarity breaking in the $L_x$-$T$ and $L_x$-$\sigma$ relations
between groups and clusters as claimed in previous work,
although the best fit $\sigma$-$T$ relations remain roughly the same
in both systems: $\sigma\propto T^{0.64}$. Alternatively,  
the significant disagreement between the observationally
fitted $L_x$-$T$ and $L_x$-$\sigma$ relations for groups and 
those expected from a perfect hydrostatic equilibrium hypothesis 
indicates that the X-ray emission of individual galaxies and 
the non-gravitational heating 
must play a potentially important role in the dynamical  
evolution of groups. Therefore, reasonable caution should be exercised in
the cosmological applications of the dynamical properties of groups.
\end{abstract}


\section{Introduction}

It has been well established that there exists a strong correlation
between the X-ray determined bolometric luminosity $L_x$,  
the X-ray temperature $T$ of the intracluster gas 
and the optical measured velocity dispersion
$\sigma$ of the cluster galaxies (Wu, Xue \& Fang 1999; hereafter Paper I
and references therein). 
A precise determination of these correlations is important 
not only for the study of dynamical properties and evolution of clusters
themselves but also for distinguishing different cosmological models 
including the prevailing estimate of the cosmic density parameter $\Omega_M$ 
through a combination of the baryon fraction  of clusters and 
the Big Bang Nucleosynthesis (White et al. 1993;
David, Jones \& Forman 1995). 
For instance, if the observed $L_x$-$T$ relation follows
$L_x\propto T^{2}$ (e.g. Quintana \& Melnick 1982; Markevitch et al. 1998), 
we would arrive at the conclusion that the X-ray 
emission is primarily due to purely gravitational heating and 
thermal bremsstrahlung, and the baryon fraction $f_b$  of clusters can 
be representative of the Universe,
i.e., $f_b$ provides  a robust estimate of $\Omega_M$ (see Paper I). 
However, if the observed $L_x$-$T$ relation appears to be $L_x\propto T^{3}$
(e.g. White, Jones \& Forman 1997), other non-gravitational heating 
mechanisms must be invoked in order to give gas sufficient 
excess energy (e.g. Ponman, Cannon \& Navarro 1999; 
Wu, Fabian \& Nulsen 1999; Loewenstein 1999), unless  
the baryon fraction of clusters is requited to vary with
temperature (David et al. 1993). The latter challenges the standard model of
structure formation.

Another critical issue is whether the $L_x$-$T$, $L_x$-$\sigma$ and 
$\sigma$-$T$ relations for clusters of galaxies can naturally extend 
to group scales. In the hierarchical model of structure formation,
groups are believed to be the scaled-down 
version of clusters, and the underlying gravitational potentials of groups and
clusters are similar when scaled to their virial radii (e.g. Navarro,
Frenk \& White 1995). It is expected that groups and clusters should 
exhibit similar correlations between $L_x$, $T$ and $\sigma$, 
provided that gas and galaxies 
are in hydrostatic equilibrium with the underlying gravitational potential
of groups and their X-ray emissions are produced by thermal bremsstrahlung.
Indeed, in the present Universe  a majority of the baryons may be bound in the 
gravitational wells of groups  (Fukugita, Hogan \& Peebles 1998). 
All groups may contain hot X-ray emitting gas with 
$kT$ around or less than $1$ keV (e.g. Price et al. 1991; Ponman et al. 1996),
and most of them should be detectable with future sensitive observations. 
Without the knowledge about the dynamical properties of groups 
characterized by the $L_x$-$T$, $L_x$-$\sigma$ and $\sigma$-$T$ relations, 
it could be misleading if the presently estimated baryon fraction of groups 
is immediately used as a cosmological indicator.
Meanwhile, any difference in these correlations between 
clusters and groups will be helpful for
our understanding of the formation and evolution of structures on scales of 
$1$--$10$ Mpc and of the significance of non-gravitational 
heating mechanisms.

The pioneering work of constructing  the  $L_x$-$\sigma$ relation for groups
was carried out by Dell'Antonio, Geller \& Fabricant (1994). 
They found a significant flattening in the relation for groups with
$\sigma$ below $300$ km s$^{-1}$, as compared with the same
relation for rich clusters. 
An extensive study of the $L_x$-$T$, $L_x$-$\sigma$ 
and $\sigma$-$T$ relations for groups was soon made by Ponman et al.(1996), 
based on 22 Hickson's compact groups (HCG) whose X-ray emissions are detected.
The most remarkable result is the steepening ($L_x\propto T^{8.2}$)
of the $L_x$-$T$ relation, in contrast with the X-ray properties of
clusters ($L_x\propto T^{2.5}$--$T^{3}$), while the significant flattening
in the  $L_x$-$\sigma$ relations for groups claimed by 
Dell'Antonio et al. (1994)
was only marginally detected. Using the RASSCALS group catalog, 
Mahdavi et al. (1997, 2000) obtained a much shallower $L_x$-$\sigma$ 
relation ($L_x\propto \sigma^{0.37\pm0.3}$)
for groups with $\sigma<340$ km s$^{-1}$ than the above results. 
Interestingly, Mulchaey \& Zabludoff (1998)
studied 12 poor ROSAT groups but arrived at an opposite conclusion that  
the X-ray component in groups follows the same $L_x$-$T$-$\sigma$ relations
as those in clusters.
Namely, groups are indeed low-mass versions of clusters.

Theoretically, several investigations have been made on the possible reasons
why the $L_x$-$T$ relation flattens with the increase of 
scale and temperature (e.g. Cavaliere, Menzi \& Tozzi 1997, 1999; 
Valageas \& Schaeffer 1999). 
In particular, if the $L_x$-$T$ and $L_x$-$\sigma$ relations show
a significant departure from the expectations of 
$L_x\propto T^{2}\propto \sigma^{4}$   
within the framework of thermal bremsstrahlung and 
hydrostatic equilibrium, the previous estimate of the amount of baryonic 
matter in groups and its application to the determination of 
cosmic density parameter would become questionable.
Consequently, one may need to study
whether the observed X-ray emission of groups has partially arisen from
the individual galaxies ( Dell'Antonio et al. 1994) or
other non-gravitational mechanisms such as those suggested by 
the preheated ICM model 
(e.g. Ponman et al. 1999) and the shock model (Cavaliere et al. 1997).

In this paper, we would like to update and then compare  
the $L_x$-$T$, $L_x$-$\sigma$ and $\sigma$-$T$ relations for groups
and clusters, using all the available X-ray/optical 
data in literature especially the new measurements of 
poor clusters and groups. We wish to achieve a better statistical 
significance, which may essentially allow us to closely examine 
the question as to whether the similarity of these relations 
would break down on group scales. 
Throughout the paper we assume a Hubble constant of
$H_0=50$ km s$^{-1}$ Mpc$^{-1}$ and a flat cosmological model of
$\Omega_0=1$.

\section{Sample}

We follow the same strategy as in Paper I to select
groups and clusters of galaxies from literature: We include all the
groups and clusters for which 
at least two parameters among the X-ray bolometric luminosity ($L_x$) and
temperature ($T$), and the optical velocity dispersion of galaxies ($\sigma$)
are observationally determined. Essentially, we use the cluster catalog
in Paper I, which contains a total of 256 clusters. We first remove
MKW9 from the list and put it into our group catalog. We then add another
19 clusters whose temperatures or velocity dispersions have 
become available since then. 
This is mainly due to the recent temperature measurements
by White (1999). Our final cluster sample consists of 274 clusters. 
Unlike the optimistic situation for X-ray clusters, the X-ray emission 
has remained  undetectable for most groups
because of their  relatively low X-ray temperatures. 
By extensively searching the literature, we find 66 groups that meet our
criteria, which include 23 HCGs (Table 1).  
As compared with the sample used in previous similar study 
by Ponman et al. (1996),  the number of groups has almost tripled.  
We convert the observed X-ray
luminosities to bolometric luminosities in the rest frame of the groups
and clusters according to an optically thin and isothermal plasma emission
model with 30 percent solar abundance, 
in which we assume $T=6$ and $1$ keV, respectively, for the 99 clusters
and 26 groups whose temperatures remain unknown spectroscopically.

 \begin{deluxetable}{llllll}
 \tablewidth{30pc}
 \scriptsize
 \tablecaption{Group Sample}
 \tablehead{
 \colhead{Group }& \colhead{redshift} &
 \colhead{$\sigma$(km/s)} & 
 \colhead{$T$ (keV)} &  
 \colhead{$L_x$ ($10^{42}$ erg/s)} & 
 \colhead{references$^a$}  }
 \startdata
 HCG12          &0.0485&  269&$ 0.89^{+ 0.12}_{- 0.12}$&$   2.04^{+  0.40}_{-  0.34}$&1, 1, 1 \nl
 HCG15          &0.0228&  457&$ 0.44^{+ 0.08}_{- 0.08}$&$   0.63^{+  0.20}_{-  0.15}$&1, 1, 1 \nl
 HCG16          &0.0132&  135&$ 0.30^{+ 0.05}_{- 0.05}$&$   0.48^{+  0.07}_{-  0.06}$&1, 1, 1 \nl
 HCG33          &0.0260&  174&$ 0.61^{+ 0.30}_{- 0.30}$&$   0.59^{+  0.17}_{-  0.13}$&1, 1, 1 \nl
 HCG35          &0.0542&  347&$ 0.91^{+ 0.18}_{- 0.18}$&$   2.23^{+  0.65}_{-  0.50}$&1, 1, 1 \nl
 HCG37          &0.0223&  $445^{+80}_{-52}$ &$ 0.67^{+ 0.11}_{- 0.11}$&$   1.32^{+  0.20}_{-  0.17}$&17, 1, 1 \nl
 HCG42          &0.0133&$  211^{+  38}_{-  34}$&$ 0.90^{+ 0.10}_{- 0.02}$&$   1.50^{+  0.06}_{-  0.07}$&2, 3, 1 \nl
 HCG48          &0.0094&  355&$ 1.09^{+ 0.21}_{- 0.21}$&$   0.38^{+  0.14}_{-  0.10}$&1, 1, 1 \nl
 HCG51          &0.0258&  263&$ 1.24^{+ 0.04}_{- 0.08}$&$   9.77^{+  2.82}_{-  2.18}$&1, 3, 1 \nl
 HCG57          &0.0304&$  344^{+  57}_{-  14}$&$ 1.15^{+ 0.21}_{- 0.08}$&$   0.95^{+  0.60}_{-  0.36}$&1, 3, 1 \nl
 HCG58          &0.0207&  178&$ 0.64^{+ 0.19}_{- 0.20}$&$   0.78^{+  0.22}_{-  0.18}$&1, 1, 1 \nl
 HCG62          &0.0137&$  376^{+  52}_{-  46}$&$ 0.95^{+ 0.03}_{- 0.03}$&$  11.00^{+  0.79}_{-  1.00}$&2, 3, 1 \nl
 HCG67          &0.0245&  240&$ 0.82^{+ 0.19}_{- 0.19}$&$   0.49^{+  0.13}_{-  0.10}$&1, 1, 1 \nl
 HCG68          &0.0080&$  183^{+  77}_{-  36}$&$ 0.54^{+ 0.15}_{- 0.15}$&$   0.19^{+  0.15}_{-  0.09}$&1, 4, 1 \nl
 HCG73          &0.0449&    96 & & $    2.69 ^{+  0.26}_{-  1.14}$&1, ..., 1 \nl
 HCG82          &0.0362&   708 & & $    1.95 ^{+  0.74}_{-  0.54}$&1, ..., 1 \nl
 HCG83          &0.0531&   501 & & $    6.46 ^{+  2.05}_{-  1.56}$&1, ..., 1 \nl
 HCG85          &0.0393&   417 & & $    1.86 ^{+  0.48}_{-  0.48}$&1, ..., 1 \nl
 HCG86          &0.0199&   302 & & $    2.09 ^{+  0.79}_{-  0.58}$&1, ..., 1 \nl
 HCG90          &0.0088&$  193^{+  36}_{-  33}$&$ 0.70^{+ 0.12}_{- 0.12}$&$   0.30^{+  0.07}_{-  0.05}$&2, 1, 1 \nl
 HCG92          &0.0215&  389&$ 0.76^{+ 0.05}_{- 0.05}$&$   1.45^{+  0.13}_{-  0.13}$&3, 3, 1 \nl
 HCG94          &0.0395&&$ 3.70^{+ 0.60}_{- 0.60}$&  114.89  & ..., 4, 4 \nl
 HCG97          &0.0218&  407&$ 0.87^{+ 0.05}_{- 0.05}$&$   6.03^{+  0.28}_{-  0.28}$&1, 1, 1 \nl
 MKW9           &0.0397&  336&$ 2.23^{+ 0.13}_{- 0.13}$&  10.50&5, 6, 5 \nl
 N34-175        &0.0283&$  589^{+ 440}_{-  31}$& & $  25.80^{+  2.76}_{-  2.76}$&8, ..., 7 \nl
 N45-384        &0.0266&$  238^{+ 125}_{-  14}$& & $   2.80^{+  0.76}_{-  0.76}$&8, ..., 7 \nl
 N56-381        &0.0295&$  265^{+  22}_{-  22}$& &     3.03 &8, ..., 9 \nl
 N56-393        &0.0221&$  422^{+  99}_{-  24}$& & $   2.56^{+  0.60}_{-  0.60}$&8, ..., 7 \nl
 N56-394        &0.0289&$  373^{+ 165}_{-  76}$& & $   3.88^{+  1.00}_{-  1.00}$&10, ..., 7 \nl
 N67-309        &0.0265&$  292^{+  97}_{-  27}$& & $   2.88^{+  0.88}_{-  0.88}$&8, ..., 7 \nl
 N67-335        &0.0204&$  471^{+  73}_{-  51}$& & $  19.80^{+  2.32}_{-  2.32}$&10, ..., 7 \nl
 N79-284        &0.0246&$  598^{+ 277}_{-  69}$& & $  1.60^{+  0.60}_{-  0.60}$     &8, ..., 7 \nl
 N79-296        &0.0232&$  359^{+  69}_{-  46}$& & $   7.68^{+  1.52}_{-  1.52}$&10, ..., 7 \nl
 N79-299A       &0.0235&$  419^{+  84}_{-  49}$& & $   5.76^{+  0.96}_{-  0.96}$&8, ..., 7 \nl
 N79-299B       &0.0235&$  384^{+ 151}_{- 110}$& & $   1.64^{+  0.56}_{-  0.56}$&8, ..., 7 \nl
 NGC383         &0.0170&  466&$ 1.50^{+ 0.10}_{- 0.10}$&  20.59&11, 11, 11 \nl
 NGC533         &0.0181&$  464^{+  58}_{-  52}$&$ 1.00^{+ 0.07}_{- 0.04}$&$   3.25^{+  0.30}_{-  0.30}$& 2, 12, 12 \nl
 NGC741         &0.0185&$  432^{+  50}_{-  46}$&$ 1.17^{+ 0.51}_{- 0.17}$&$   3.03^{+  0.98}_{-  0.98}$& 2, 12, 12 \nl
 NGC2300        &0.0076&  251&$ 0.88^{+ 0.04}_{- 0.05}$&  44.90&3, 3, 13 \nl
 NGC2563        &0.0163&$  336^{+  44}_{-  40}$&$ 1.39^{+ 0.05}_{- 0.06}$&$   1.70^{+  0.20}_{-  0.20}$ & 2, 3, 12 \nl
 NGC3258        &0.0095&  400&$ 1.85^{+ 0.23}_{- 0.22}$&$  17.20^{+  1.20}_{-  1.20}$&14, 3, 14 \nl
 NGC4104        &0.0283&  546&$ 2.16^{+ 0.15}_{- 0.18}$& & 15, 3, ... \nl
 NGC4325        &0.0252&$  265^{+  50}_{-  44}$&$ 1.05^{+ 0.04}_{- 0.03}$&$  10.05^{+  2.77}_{-  2.77}$ & 2, 3, 12 \nl
 NGC5044        &0.0082&  360&$ 1.00^{+ 0.02}_{- 0.02}$&  17.60&3, 3, 16 \nl
 NGC5129        &0.0237&$  294^{+  43}_{-  38}$&$ 0.87^{+ 0.02}_{- 0.05}$&$   5.65^{+  2.08}_{-  2.08}$& 2, 3, 12 \nl
 NGC5846        &0.0063&$  368^{+  72}_{-  61}$&$ 0.65^{+ 0.13}_{- 0.09}$&$   9.60^{+  1.99}_{-  1.99}$&2, 12, 12 \nl
 RGH12          &0.0273&$  179^{+  80}_{-  36}$& & $   2.10^{+  0.80}_{-  0.80}$&17, ... 17 \nl
 RGH48          &0.0107&$  105^{+  24}_{-  16}$& & $   1.20^{+  0.20}_{-  0.20}$&17, ... 17 \nl
 RGH73          &0.0231&$  685^{+  85}_{-  62}$& & $   4.60^{+  1.30}_{-  1.30}$&17, ... 17 \nl
 RGH80          &0.0367&$  431^{+ 251}_{-  92}$&$ 1.04^{+ 0.02}_{- 0.05}$&$  15.60^{+  3.10}_{-  3.10}$&17, 3, 17 \nl
 RGH85          &0.0246&$  267^{+  74}_{-  41}$& & $   4.30^{+  0.80}_{-  0.80}$&17, ..., 17 \nl
 RXJ1340.6+4019 &0.1710&$  380^{+ 350}_{- 110}$&$ 0.92^{+ 0.08}_{- 0.08}$&  45.00&18, 18, 18 \nl
 RXJ1724.1+7000 &0.0378&&$ 0.95^{+ 0.20}_{- 0.20}$&    5.28  & ..., 19, 19 \nl
 RXJ1736.4+6804 &0.0258&  288&$ 1.12^{+ 0.40}_{- 0.20}$&  11.40&19, 19, 19 \nl
 RXJ1751.2+6532 &0.0386&&$ 1.25^{+ 0.25}_{- 0.10}$&    5.13  & ..., 19, 19 \nl
 RXJ1755.8+6236 &0.0258&  387&$ 1.50^{+ 0.50}_{- 0.20}$&  12.33&19, 19, 19 \nl
 RXJ1756.5+6512 &0.0258&  195&$ 0.80^{+ 0.05}_{- 0.05}$&   4.49&19, 19, 19 \nl
 RXJ1833.6+6520 &0.0370&&$ 0.92^{+ 0.10}_{- 0.10}$&    4.81  & ..., 19, 19 \nl
 Pegasus        &0.0140&  780&$ 1.05^{+ 0.02}_{- 0.03}$& & 20, 3, ... \nl
 S34-111        &0.0173&$  486^{+  53}_{-  37}$& & $   6.76^{+  0.92}_{-  0.92}$&8, ..., 7 \nl
 S34-113        &0.0172&$  644^{+  62}_{-  49}$&$ 1.60^{+ 0.50}_{- 0.50}$&   21.71  & 10, 9, 9 \nl
 S34-115        &0.0225&$  474^{+ 168}_{-  93}$& & $   5.20^{+  0.96}_{-  0.96}$&8, ..., 7 \nl
 S49-140        &0.0179&$  205^{+  59}_{-  21}$& & $   3.60^{+  0.64}_{-  0.64}$&8, ..., 7 \nl
 S49-142        &0.0211&$   69^{+   2}_{-   2}$& & $   1.80^{+  0.60}_{-  0.60}$&8, ..., 7 \nl
 S49-146        &0.0250&$  617^{+ 132}_{- 110}$& & $   4.08^{+  0.88}_{-  0.88}$&8, ..., 7 \nl
 S49-147        &0.0191&$  233^{+ 141}_{-  43}$& & $   4.96^{+  0.84}_{-  0.84}$&8, ..., 7 \nl
\tablenotetext{a}{Listed in order, are sources for velocity dispersion, 
temperature and X-ray luminosity, respectively: 
(1)Ponman et al. 1996; 
(2)Zabludoff \& Mulchaey 1998;
(3)Davis, Mulchaey \& Mushotzky 1999;
(4)Pildis, Bregman \& Evrard 1995;
(5)Buote 1999;
(6)Horner, Mushotzky \& Scharf 1999;
(7)Burns et al. 1996;
(8)Ledlow et al. 1996;
(9)Price et al. 1991;
(10)Dell'Antonio, Geller \& Fabricant 1994;
(11)Komossa \& B\"ohringer 1999;
(12)Mulchaey \& Zabludoff 1998; 
(13)Mulchaey et al. 1993;
(14)Pedersen, Yoshii \& Sommer-Larsen 1997; 
(15)Beers et al. 1995;
(16)David et al. 1994;
(17)Mahdavi et al. 1997;
(18)Jones, Ponman \& Forbes 1999;
(19)Henry et al. 1995;
(20)Fadda et al. 1996.
}
 \enddata
 \end{deluxetable}

\section{Results}

Essentially, two linear regression methods are used in the fit of 
the observed data ($L$, $T$ and $\sigma$) to a power-law relation: the standard
ordinary least-square (OLS) method and the orthogonal distance regression
(ODR) technique (Feigelson \& Babu 1992). The latter is preferred because
it can account for data scatters around two variables, which is 
particularly suited when two variables contain significant measurement
uncertainties. We perform both the OLS and ODR fits to the data sets of 
our group and cluster samples, respectively. 
In order to maximally use the published data especially 
for groups in the ODR fitting, we assign the average values of
measurement uncertainties in $L_x$, $T$ and $\sigma$ to those 
data whose error bars are not available. Specifically, 
the average relative errors ($\Delta L_x/L_x$, $\Delta T/T$,
$\Delta\sigma/\sigma$) are found to be ($14.7\%$, $22.4\%$, $16.5\%$)  
and ($24.7\%$, $15.5\%$, $28.4\%$) for clusters and groups, respectively.
Finally, the Monte-Carlo simulations are performed to estimate the 
standard deviations in the fitted relations.

\subsection{$L_x$-$T$ relation}

The observed and our best-fit $L_x$-$T$ relations are shown in Fig.1
and also summarized in Table 2. It appears that the resultant 
$L_x$-$T$ relation for (184) clusters remains nearly the same as that given
in Paper I for 142 clusters, $L_x\propto T^{2.79\pm0.08}$.
However, our best fit  
$L_x$-$T$ relation for 38 groups becomes somewhat flatter: The power-index 
in the ODR fitting drops to $5.57\pm1.79$, in contrast to the value of
$8.2\pm2.7$ reported by Ponman et al. (1996) based on 16 HCGs.
Nevertheless, at $3\sigma$ significance level we have confirmed 
their claim for the similarity breaking of the $L_x$-$T$ relation
at the low temperature end.

 \begin{deluxetable}{ccllll}
 \tablewidth{35pc}
 \scriptsize
 \tablecaption{The best fit $L_x$-$T$ relations$^a$}
 \tablehead{
 \colhead{sample}& \colhead{number} &
 \colhead{OLS} & 
 \colhead{ODR} &
 \colhead{$\tau$} &
 \colhead{$P$} }
 \startdata
 group   & 38  &    $L_x=10^{-0.39\pm0.19}T^{2.10\pm0.10}$ &
                    $L_x=10^{-0.27\pm0.05}T^{5.57\pm1.79}$ 
               &    $0.460$ &  $4.852\times10^{-5}$ \nl
cluster  & 184 &    $L_x=10^{-0.89\pm0.08}T^{2.54\pm0.11}$ &
                    $L_x=10^{-0.032\pm0.065}T^{2.79\pm0.08}$ 
               &    $0.686$ &  $1.569\times10^{-43}$ \nl
combined & 222 &    $L_x=10^{-0.17\pm0.19}T^{2.85\pm0.04}$ &
                    $L_x=10^{-0.14\pm0.05}T^{3.03\pm0.06}$ 
               &    $0.763$ &  $0$ \nl
\tablenotetext{a}{$L_x$ and $T$ are in units of $10^{42}$ erg s$^{-1}$ and keV,
	   respectively.}
 \enddata
 \end{deluxetable}

\placefigure{fig1}

\subsection{$L_x$-$\sigma$ relation}

We display in Fig.2 and Table 3 the X-ray luminosity-velocity dispersion
relations for 59 groups and 197 clusters. Again, the $L_x$-$\sigma$
relation for clusters agrees with our previous result based on 156 clusters
(Paper I): $L_x\propto\sigma^{5.3}$, while the best fit 
$L_x$-$\sigma$ relation for our group sample  
reads $L_x\propto \sigma^{2.35\pm0.21}$, which
is significantly flatter than both the above relation for clusters and 
the Ponman et al. (1996) result for 22 HCGs 
$L_x\propto\sigma^{4.9\pm2.1}$. 
Yet, our $L_x$-$\sigma$ relation for groups has not reached 
the shallow slopes (0.37 -- 1.56) reported by Mahdavi et al. (1997, 2000).
A glimpse of Fig.2 reveals the following two features: (1)The data of 
groups and clusters are joined together through poor cluster 
population, and there is no apparent gap in between; 
(2)The scatters of $\sigma$ around the best fit $L_x$-$\sigma$ relation
become relatively large on group scale.

 \begin{deluxetable}{ccllll}
 \tablewidth{35pc}
 \scriptsize
 \tablecaption{The best fit $L_x$-$\sigma$ relations$^a$}
 \tablehead{
 \colhead{sample}& \colhead{number} &
 \colhead{OLS} & 
 \colhead{ODR} &
 \colhead{$\tau$} &
 \colhead{$P$} }
 \startdata
 group   & 59  &    $L_x=10^{-2.95\pm0.30}\sigma^{1.00\pm0.12}$ &
                    $L_x=10^{-6.38\pm0.55}\sigma^{2.35\pm0.21}$ 
               &    $0.280$ &  $1.720\times10^{-3}$ \nl
cluster  & 197 &    $L_x=10^{-7.18\pm0.55}\sigma^{2.72\pm0.19}$ &
                    $L_x=10^{-13.68\pm0.61}\sigma^{5.30\pm0.21}$
               &    $0.527$ &  $3.783\times10^{-28}$ \nl 
combined & 256 &    $L_x=10^{-8.75\pm0.25}\sigma^{3.55\pm0.06}$ &
                    $L_x=10^{-12.10\pm0.52}\sigma^{4.75\pm0.18}$
               &    $0.634$ &  $0$ \nl 
\tablenotetext{a}{$L_x$ and $\sigma$ are in units of $10^{42}$ erg s$^{-1}$
           and km s$^{-1}$, respectively.}
 \enddata
 \end{deluxetable}

\placefigure{fig2}

\subsection{$\sigma$-$T$ relation}

The best fit $\sigma$-$T$ relation for clusters using 109 clusters
remains almost unchanged (Fig.3 and Table 4) as compared with 
the previous studies by  Wu, Fang \& Xu (1998; and references therein)
and Paper I: $\sigma\propto T^{0.65\pm0.03}$. Meanwhile, we have found
roughly the same relation for our  sample of 36 groups:
$\sigma\propto T^{0.64\pm0.08}$, which is also consistent with the previous
results within uncertainties (Ponman et al. 1996; Mulchaey \& Zabludoff 1998). 
For clusters, this relation alone indicates
that the intracluster gas may not be in isothermal and hydrostatic
equilibrium with the underlying gravitational potential of clusters.
However, the same conclusion is not strictly applicable to groups 
if the large uncertainty in  the presently fitted $\sigma$-$T$ relation 
is included.  Additionally, 
the ratios of specific energy in galaxies to that in gas for the 36
groups exhibit rather large scatters ranging from 0.3 to 3.5, with
an average value of $\beta_{spec}=0.85$.

 \begin{deluxetable}{ccllll}
 \tablewidth{35pc}
 \scriptsize
 \tablecaption{The best fit $\sigma$-$T$ relations$^a$}
 \tablehead{
 \colhead{sample}& \colhead{number} &
 \colhead{OLS} & 
 \colhead{ODR} & 
 \colhead{$\tau$} &
 \colhead{$P$} }
 \startdata
 group   & 36  &    $\sigma=10^{2.51\pm0.19}T^{0.45\pm0.07}$ &
                    $\sigma=10^{2.53\pm0.01}T^{0.64\pm0.08}$ 
               &    $0.355$ &  $2.291\times10^{-3}$ \nl 
cluster  & 109 &    $\sigma=10^{2.53\pm0.03}T^{0.58\pm0.05}$ &
                    $\sigma=10^{2.49\pm0.02}T^{0.65\pm0.03}$
               &    $0.583$ &  $2.535\times10^{-19}$ \nl  
combined & 145 &    $\sigma=10^{2.53\pm0.19}T^{0.57\pm0.01}$ &
                    $\sigma=10^{2.51\pm0.01}T^{0.61\pm0.01}$
               &    $0.703$ &  $4.135\times10^{-36}$ \nl  
\tablenotetext{a}{$\sigma$ and $T$ are in units of km s$^{-1}$ and keV,
	   respectively.}
 \enddata
 \end{deluxetable}

\placefigure{fig3}

\subsection{Co-consistency test}

The employment of the ODR fitting method essentially ensures that 
the best fit relations  between $L_x$, $T$ and $\sigma$ are self-consistent
(Paper I).  We now examine the co-consistency between 
these relations in the sense that these three relations are not independent.
Our strategy is as follows: We first derive the correlation between 
$\sigma$ and $T$ from the best fit $L_x$-$T$ and 
$L_x$-$\sigma$ relations listed in Table 2 and Table 3. We then
compare this `derived' $\sigma$-$T$ relation with the one 
obtained independently from our ODR fitting shown in Table 4. 
Our fitted relations should be acceptable if a good agreement between
the derived and fitted $\sigma$-$T$ relations is reached. Otherwise,
at least one of the three fitted relations will be questionable.

For cluster sample, combining the $L_x$-$T$ and $L_x$-$\sigma$ relations 
in Table 2 and Table 3 yields $\sigma\propto T^{0.53\pm0.04}$.
Therefore, at $2\sigma$ significance level the derived $\sigma$-$T$ relation 
is consistent with the directly fitted one $\sigma\propto T^{0.65\pm0.03}$.
As for the group sample,  the derived $\sigma$-$T$ relation reads
$\sigma\propto T^{2.37\pm0.90}$, which compares with 
the best fit one $\sigma\propto T^{0.64\pm0.08}$.
Regardless of the apparent disagreement between the mean slopes,  
the large $68\%$ confidence interval makes the two relations 
show no difference within $2\sigma$.
As a result, the three fitted relations for groups 
in Table 2 -- Table 4 are still consistent with each other 
when the $2\sigma$ uncertainties are taken into account.
Indeed, a visual examination of Fig.1--Fig.3 reveals that  
the data points of groups show very 
large dispersions especially on the $L_x$-$\sigma$ plane.
Either the sparse data points and/or the inclusion of a few unusual groups
in our fitting (e.g. HCG94, S49-142, etc.) may account for the 
the marginal co-consistency between the $L_x$-$T$, $L_x$-$\sigma$ and 
$\sigma$-$T$ relations for groups.

In Table 2 -- Table 3, we have also given the significance level 
for the resulting correlation coefficient for each data set, 
based on Kendall's $\tau$.  It appears that the probability of
erroneously rejecting the null hypothesis of no correlation
between $L_x$, $T$ and $\sigma$ is $P<0.2\%$ for all the cases. 
Therefore, it is unlikely that the correlations we have found 
for groups and clusters are an artifact of the small samples
and/or the statistical fluke.

\section{Discussion and conclusions}

\subsection{Global properties}

Groups and clusters constitute large and nearly virialized dynamical systems
in the Universe.  While the distribution of dark matter  in these systems
may exhibit a regularity such as the universal density profile,
the hot intragroup/intracluster gas could be affected by non-gravitational 
heating mechanisms (e.g. star formation), especially for poor
systems like groups of galaxies where even individual galaxies may 
make a nonnegligible contribution to the X-ray emission 
(Dell'Antonio et al.  1994; Mulchaey \& Zabludoff 1998).  
Therefore, the standard scenario that groups and clusters should be 
the similar dynamical systems at different mass ranges is only applicable 
to the distribution and evolution of the dark matter particles.
Whether or not the hot gas can be used as a good tracer of  the underlying
gravitational potential wells depends on how significant 
the X-ray emission of individual galaxies and the non-gravitational
heating would be, which in turn depends on how massive a dynamical system 
will be.  
Because groups of galaxies connect individual galaxies to clusters of 
galaxies, one may expect to detect the transition between 
``galaxy-dominated'' and ``intracluster medium dominated'' properties 
occurring on group scales (e.g. Dell'Antonio et al.  1994).

Such a scenario has essentially been justified by  
our analysis of the correlations between X-ray luminosity, temperature 
and velocity dispersion for  groups and clusters, 
although the current data especially for groups still have rather 
large uncertainties. It is remarkable that 
we have detected the similarity breaking in the $L_x$-$T$ and
$L_x$-$\sigma$ relations between massive systems (clusters) and low mass
ones (groups), indicating that the overall dominant X-ray emission and
heating mechanisms
are quite different in these two systems. Nevertheless, 
for the $\sigma$-$T$ relation we have not found an apparent discrepancy
between the two systems. Our result essentially agrees  
with the previous findings of Dell'Antonio et al. (1994), 
Ponman et al. (1996) and Mahdavi et al. (1997, 2000).

\subsection{Uncertainties}

Major uncertainties in the presently fitted $L_x$-$T$, 
$L_x$-$\sigma$ and $\sigma$-$T$ relations for groups and clusters
come from the sparse (X-ray) data sets of poor clusters and groups.
Although an observational determination of these relations does not 
in principle require the completeness of the sample, the reliability and
significance of our fitting can be seriously affected by the small number
statistics. The marginal co-consistency between the three
relations for groups may be partially due to the small group sample.
It is generally believed that poor clusters and groups 
should all contain  hot X-ray emitting gas 
(e.g. Price et al. 1991; Ponman et al. 1996).
However, most of them can be hardly detected by current X-ray observations 
because of their low temperatures $T\sim0.1$--$1$ keV. Therefore, 
our group sample suffers from a selection bias, in which the present 
X-ray data are acquired by different authors with different criteria. 
For instance, about 1/3 of our group sample are HCGs, 
and the steepening of the $L_x$-$T$ relation for groups may have partially 
arisen from the contribution of these HCG populations.

Our statistical results are sensitive to the linear fitting methods, especially
for the $L_x$-$T$ and $L_x$-$\sigma$ relations. This arises because 
the observed quantities, $L_x$, $T$ and $\sigma$, all have measurement
uncertainties, while OLS method ignores scatters around the horizontal
variable (e.g. $T$ or $\sigma$). At this point, the ODR fitting technique is 
strongly recommended.  However, not all the data points have information 
about their measurement uncertainties, or some error bars are difficult 
to evaluate. In this case, we have simplified the problem and employed 
the average values instead, in order to maximally use the available data 
points, which may have yielded further uncertainties in the ODR fitted 
relations. Note that the measurement uncertainties  in $L_x$ are 
relatively smaller than those in $T$ and $\sigma$ 
(see Fig.1 and Fig.2). This explains the 
significant difference in the resultant $L_x$-$T$ and $L_x$-$\sigma$ 
relations between  OLS and ODR methods.

\subsection{Physical implications}

Under the assumption that both gas and galaxies are in hydrostatic
equilibrium with the underlying gravitational potential of 
group/cluster, we would expect that the total X-ray luminosity via
thermal bremsstrahlung scales as (e.g. Paper I)
\begin{equation}
L_x\propto T^{2.5}f_b^2 r_{c}^{-1} S_{gas},
\end{equation}
and 
\begin{equation}
L_x\propto \sigma^{4} T^{1/2} f_b^2 r_{c}^{-1} S_{gal},
\end{equation}
where $f_b$ is the (gas) baryon fraction, $r_c$ is the core radius and 
$S$ is the so-called dimensionless structure factor which depends 
strongly on the power-index of gas/galaxy profile (e.g. the conventional
$\beta$ model or the King model) but weakly on the core radius.
Using the emission weighted temperature instead of $T$ in eq.(1) only 
leads to a modification to $S_{gas}$. Additionally, the velocity dispersion
of galaxies and the temperature of hot X-ray emitting gas satisfy
\begin{equation}
\sigma \propto  T^{1/2}. 
\end{equation}

From our fitted $\sigma$-$T$ relations alone,  groups are still consistent with
$\sigma \propto  T^{1/2}$ within $2\sigma$ uncertainties, 
while a significant deviation from what is expected under 
the scenario of isothermal and hydrostatic equilibrium is found for clusters.
The latter is consistent with a number of similar studies on the issue 
(e.g. White, Jones \& Forman 1997; Wu et al. 1998; 1999).
Yet, it is unlikely that the intragroup gas is in the state of a perfect 
isothermal and hydrostatic equilibrium with the underlying gravitational
potential as allowed by the $\sigma$-$T$ relation within its $95\%$ 
significance interval.
This point can be further demonstrated by the apparent disagreement between 
the theoretical prediction of eq.(1), $L_x\propto T^2$, 
and our fitted $L_x$-$T$ relation  for groups in Table 2, 
$L_x\propto T^{5.57\pm1.79}$, unless the baryon fraction is 
assumed to vary with gas temperature. Meanwhile, the best fit $L_x$-$\sigma$
relation for groups, $L_x\propto \sigma^{2.35\pm0.21}$, differs from the 
theoretical expectation (eq.[2]) in conjunction with eq.(3): 
$L_x\propto \sigma^{4}$.  Taking these results as a whole, we feel that 
the currently available optical/X-ray data have already provided
convincing evidence for either the failure of isothermal and 
hydrostatic equilibrium hypothesis for intragroup gas or 
the significant contribution of X-ray emission from the individual 
group galaxies. Therefore, 
any cosmological applications of these fitted relations without considering
these effects could be misleading.

The physical implications of the $L_x$-$T$, $L_x$-$\sigma$ and 
$\sigma$-$T$ relations for clusters have been extensively studied in
Paper I. For the group sample, there exists an apparent
difference between the theoretically predicted 
$L_x$-$T$ and $L_x$-$\sigma$ 
relations (eq.[1]) and the observationally determined ones. 
Dell'Antonio et al. (1994) attributed the flattening of $L_x$-$\sigma$ 
relation to the additional X-ray emission of individual group galaxies, while 
Ponman et al (1996) interpreted the steepening of the  $L_x$-$T$
relation as the result of the wind injection from galaxies.
The recent detection of the X-ray wakes in pool cluster A160 may give a 
strong support to these scenarios, i.e., a large fraction of 
X-ray emission in groups and poor clusters can be associated with
individual galaxies (Drake et al. 2000).
It thus deserves to  explore whether
this scenario can  quantitatively account 
for the reported $L_x$-$T$-$\sigma$ relations 
for groups. Alternatively, other mechanisms such as 
preheating by supernovae and substructure merging may
also contribute extra heating to the intragroup medium.

On the other hand, the prevailing determination of the baryon fraction of 
groups via hydrostatic equilibrium and thermal bremsstrahlung
could be seriously contaminated by the X-ray emission of individual 
group galaxies and other non-gravitational heating.
This might account for the relatively large variations of 
the presently derived baryon fractions  among different groups.
It is unlikely that a robust estimate of the baryon
fractions of groups and poor clusters within the framework of conventional 
dynamical models and their cosmic evolution can be achieved without
a better understanding of the various heating mechanisms.
Indeed, we cannot exclude the possibility that
some puzzles about the properties of the baryon fractions of groups
and clusters are due to the contamination of the X-ray emission of 
individual galaxies and the non-gravitational heating.  
For instance, the standard model predicts an increase 
in the baryon fraction with radius, and a universal value of $f_b$ 
at large radii cannot be reached unless the X-ray temperature is required
to rise in some clusters (Wu \& Xue 2000 and references therein).
The presence of such puzzle has essentially prevented the baryon fractions 
of groups and clusters from the cosmological applications.  
Inclusion of the contributions of other emission/heating sources  
in the estimate of the total
gravitational masses of groups and clusters may lead to a re-arrangement of
the radial distribution of baryon fraction.  Further work will thus be    
needed to explore whether these effects can resolve the above puzzle.

\acknowledgments
We thank the referee, Andisheh Mahdavi, for many valuable suggestions
and comments. This work was supported by 
the National Science Foundation of China, under Grant No. 19725311.

\clearpage

\figcaption{The $L_x$-$T$ relations for 184 clusters (open circles)
and 38 groups (filled circles). 
The dotted and solid lines are the best ODR fitted relations 
to the group and cluster samples, respectively.
\label{fig1}}

\figcaption{The $L_x$-$\sigma$ relations for 197 clusters 
and 59 groups. The notations are the same as in Fig.1.
\label{fig2}}

\figcaption{
The $\sigma$-$T$ relations for 109 clusters 
and 36 groups. The notations are the same as in Fig.1.
\label{fig3}}

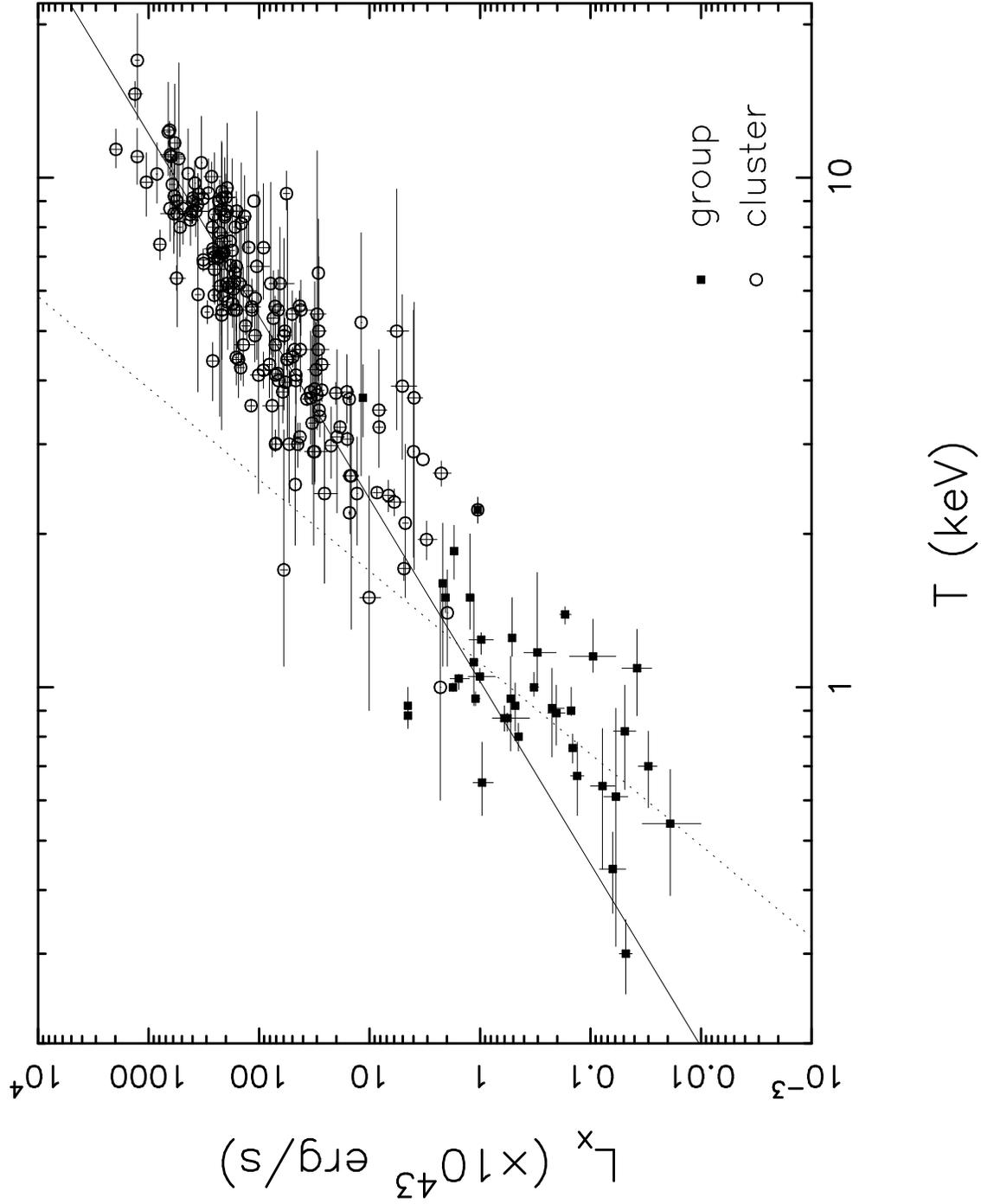
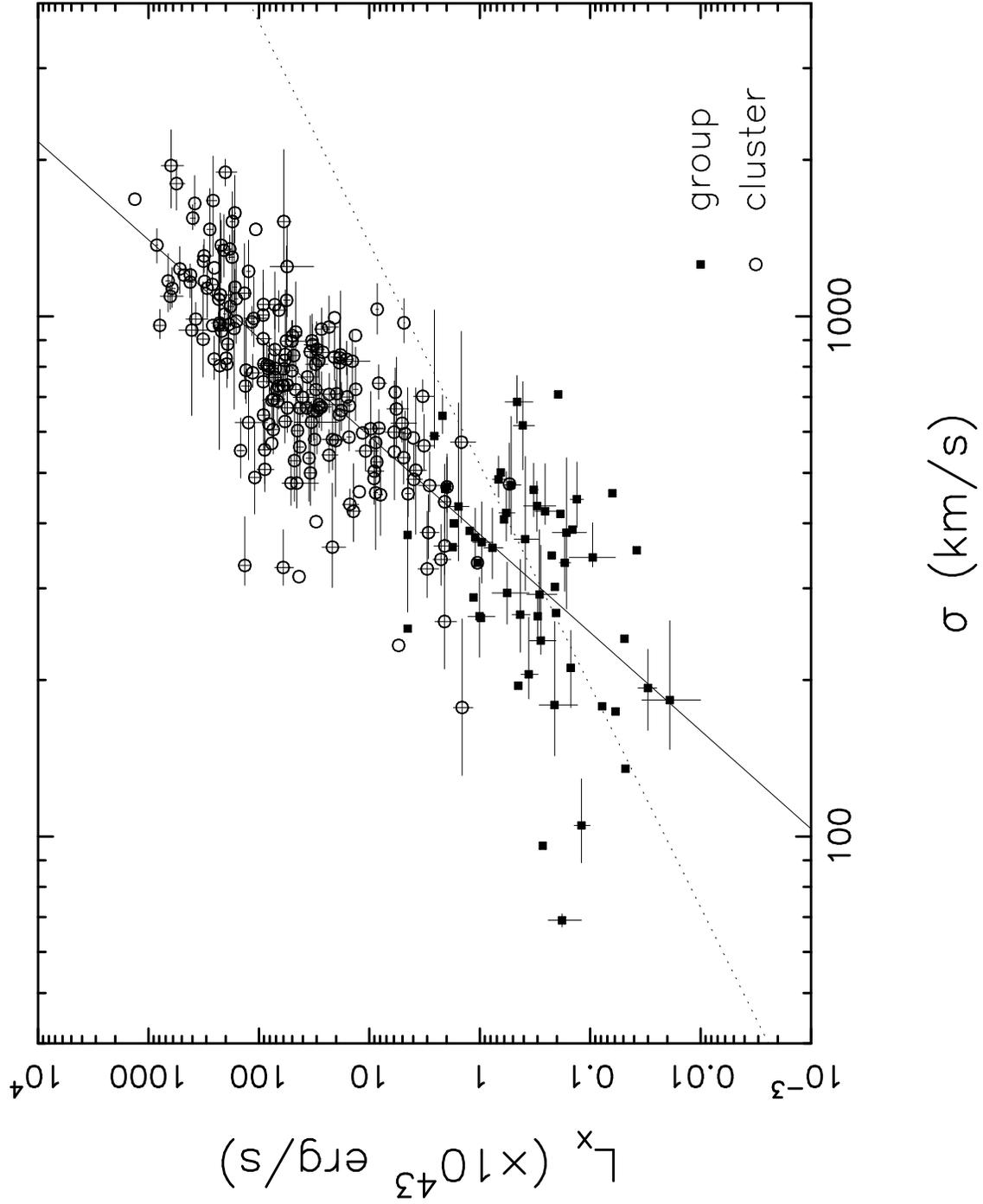
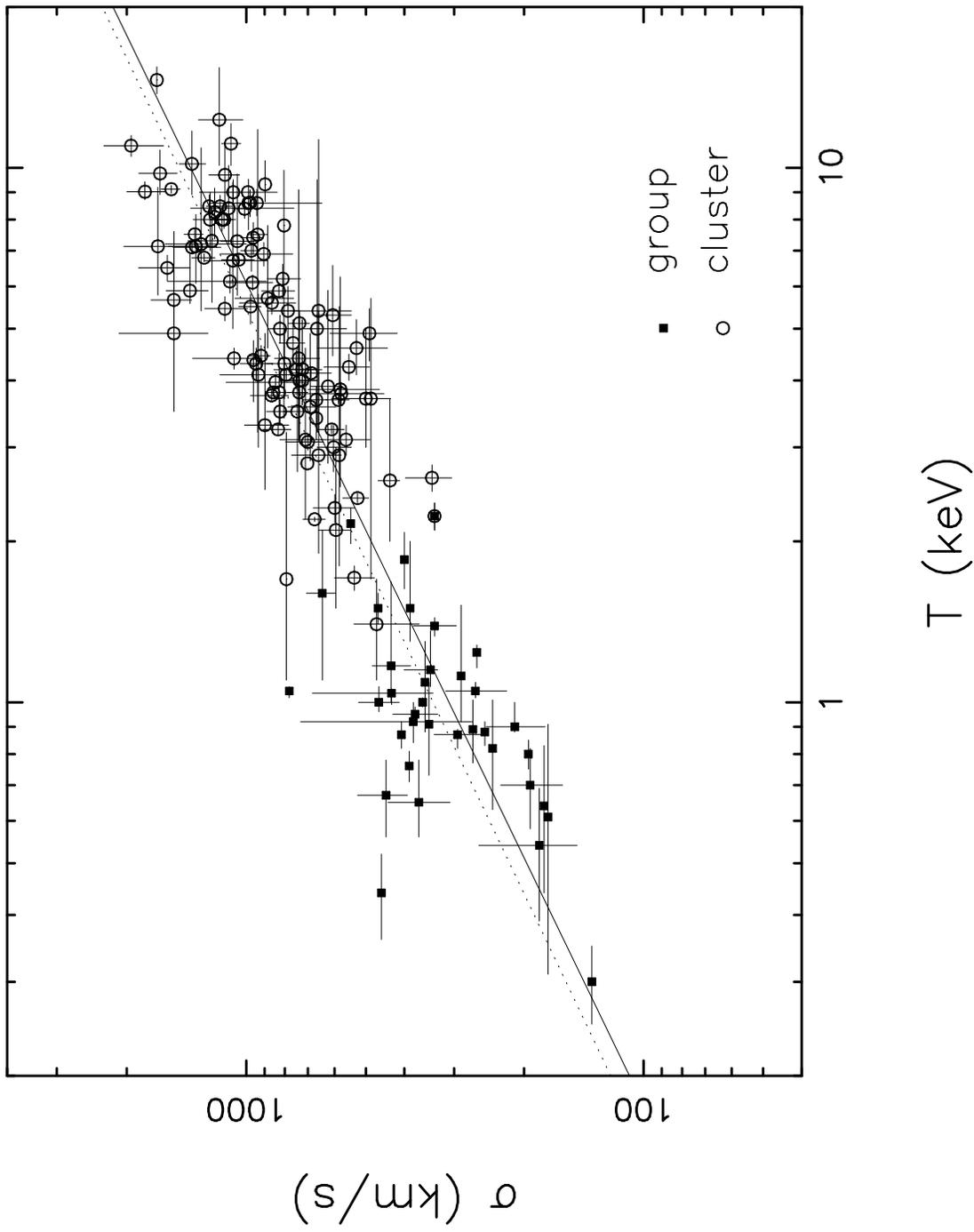
\end{document}